\documentstyle{mn}
\newcommand{\bgr}{\bibitem[\protect\citename{dummy }1893]{dum}}
\newcommand{\etal}{et~al.\ }
\newcommand{\etalc}{et~al.}
\newcommand{\eg}{e.g.\ }
\newcommand{\ie}{i.e.\ }
\newcommand{\beq}{\begin{equation}}
\newcommand{\eeq}{\end{equation}}
\newcommand{\aldo}{\alpha '}

\newcommand{\gta}{\mbox{\small\raisebox{-0.6ex}{$\,\stackrel
{\raisebox{-.2ex}{$\textstyle >$}}{\sim}\,$}}}
\begin{document}
\title[The flaring behaviour of 3C\,279]
{Modelling the millimetre--infrared flaring behaviour of the quasar
1253$-$055 (3C\,279)}

\author[S. J. Litchfield \etal]
{S.~J.~Litchfield,$^{1}$\thanks{Present address: University of Crete,
Department of Physics, 714 09 Heraklion, Crete, Greece.}
J.~A.~Stevens,$^{1}$ E.~I.~Robson,$^{1,2}$ W. K. Gear$^3$\\
$^{1}$Centre for Astrophysics, University
of Central Lancashire, Preston, PR1 2HE\\
$^{2}$Joint Astronomy Centre, 660 N. A'oh\={o}k\={u} Place, University
Park, Hilo, Hawaii 96720, USA\\
$^3$Royal Observatory, Blackford Hill, Edinburgh, EH9 3HJ
}

\date{Received; accepted}

\maketitle
\begin{abstract}
The infrared through millimetre light curve of 3C\,279 is investigated
for the period of 1986 until mid-1994, during which time several flares
were observed. A quiescent spectrum (identified with emission from an
underlying jet) is presented. Both the near-IR and 375--150 GHz regimes are
shown to be well described by power laws, with no evidence for any
thermal contribution in the IR.
Successful isolation of the flaring component by subtraction of a
base level is found to be difficult.
Dividing each individual flare into two regimes corresponding to before
and after maximum flux, we find
strong linear correlations between log 90 GHz
flux and 22--90 GHz spectral slope. Furthermore,
the gradient of the linear correlation
steepens as the flare decays after maximum. This trend is observed for
several successive flares, and can be successfully
explained in terms of evolution of the flare according to the
the shocked-jet model of Marscher \& Gear (1985).
\end{abstract}


\section{Introduction}
\begin{figure*}
\vbox to220mm{\vfil Landscape figure 1 to go here. \vfil}
\caption[dum]{Light curves for 3C\,279.}
\label{petrel}
\end{figure*}
The object 1253$-$055 (3C\,279)
is a member of that class of active galactic nucleii
(AGN) known as blazars,
comprising BL\,Lacertae objects and optically
violently variable (OVV) quasars (see Bregman 1990 for a review).
3C\,279 is a member of the latter group,
and exhibits those properties evinced by sources
typical of the blazar phenomenon: strongly variable radio to optical
polarization (\eg Hughes, Aller \& Aller 1991,
Wills \etal 1992, Nartallo-Garcia \etalc, in preparation),
significant flux variability (\eg Webb \etal 1990, Edelson 1992, Stevens
\etal 1994 [hereafter BLZ5]),
and a flat or slowly declining
radio spectrum (\eg Brown \etal 1989a, Gear \etal 1994).
In addition it is a strong superluminal radio source (Unwin \etal 1989) and has
shown highly variable emission at energies in excess of 1 GeV (Kniffen \etal
1993).

In a previous paper (BLZ5) we presented light curves for our complete
sample of sources from K-band in the near-infrared to 22 GHz, including a
representative spread of frequencies around the transition from optically
thin to thick emission (375, 270, 230, 150 GHz). The flaring behaviour
was analysed in the context of the shocked-jet model of Marscher \& Gear
(1985); hereafter MG85. In this
model the individual flares evolve through three distinct phases,
depending on the dominant energy loss mechanism. In the initial growth phase
Compton energy losses predominate and the synchrotron self-absorption
turn-over frequency decreases whilst the turn-over flux increases. The
second (synchrotron or plateau) phase occurs when synchrotron losses dominate.
In this phase the turn-over flux is roughly constant whilst the turn-over
frequency decreases.
Finally the flare reaches the adiabatic or decay phase when both the turn-over
flux and frequency decrease. The appearance of the light curve for any
given frequency depends
on whether the observing frequency lies on the rise, plateau or decay phase.
Flare maximum is attained simultaneously at observing frequencies
above the transition from Compton to synchrotron phase; the amplitude
of the flare increases with decreasing frequency. Below this transition, flare
maximum occurs when the turn-over passes the observing frequency; lags between
maxima correspond to the time taken for the turn-over to evolve to longer
wavelengths.
It was found that, for a number of sources (3C\,279
included), the delays between flare maxima at different frequencies and the
amplitudes of the flares at those frequencies were roughly in agreement
with the above picture. However, spectral deconvolution of
the flaring emission from any underlying component is necessary to confirm
the details of the model described above, and furthermore to provide
necessary information which can lead to refinement of the model. The aim
of this paper is to look at ways in which the data presented in
BLZ5 can be used to aid development of shocked-jet models.

\section{Observations}
This paper analyses substantially the same data as
was presented in BLZ5, and the observations
are discussed in detail in that paper (see also Robson \etal 1993).
To summarize, the bulk of the 150--375 GHz
data were obtained at the James Clerk Maxwell Telescope, Mauna Kea, using
the UKT14 bolometer (Duncan \etal 1990). Additional 230 and 150 GHz data,
as well as all the 90 GHz observations,
were obtained from the IRAM 30-m telescope at Pico Valeta (see
Steppe \etal 1993 for instrument and reduction details), and from
the SEST, La Silla (see Booth \etal 1989). The 37 and 22 GHz data came
from the Mets\"{a}hovi Radio Research Station, Kylm\"{a}l\"{a}
(Ter\"{a}sranta \etal 1992). The infrared observations were obtained mostly
at the United Kingdom Infrared Telescope
(UKIRT), Mauna Kea, or at the ESO 2.2-m telescope, Chile (see
Litchfield, Robson \& Stevens 1994 and references therein for data and
reduction details).

\section{Results}
\subsection{Overview of data}
The light curves for 3C\,279 from the beginning of 1986 until mid 1994
are presented in Fig.\ \ref{petrel}. The K-band is shown
as being representative of the near-IR; the full 1.25--10.0 $\mu$m data
set is tabulated in Litchfield \etal 1994.

The key features of the behaviour of the source can be summarized as follows.
There was a major flare in 1988 which was well covered at IR frequencies and at
37, 22 GHz, but largely under-sampled elsewhere. Subsequent flares were
well sampled at all frequencies except the near-IR.
Between mid 1989 and early 1990 there was a period of low
flux, followed by a large flare, the onset of which began
in early 1990. This flare has the appearance of being two flares occurring
in rapid succession: the flare of early 1990 is seen to reach a maximum in
late June at 150 and
90 GHz, and then to decline slightly before rapidly rising to a
maximum in late 1990/early 1991.  At 90 and 37 GHz the rise is clearly broken
into two regions of markedly different gradient: the rate of change of flux
increases sharply in the latter part of 1990.
This indicates the onset of a
new flare at this time, since the onset of a flare is seen to be much more
rapid than the subsequent decay (Robson \etal 1993).

The flare decreased through 1991 at all frequencies except for 22 GHz. This
decline was reversed in late 1991 with a further small peak
followed by a decline to a level concordant with quiescence
(at JCMT frequencies) in mid 1992. For frequencies below 150 GHz the low
level achieved in 1992 was much higher than the quiescent level.
This indicates the slower decay time-scales
of flares at lower frequencies which is a feature of the MG85 model.
Further small flares occurred in 1993 and 1994.
It should be noted that, at lower
frequencies (particularly 22 GHz) the individual features have merged into a
broader but shallower hump, and the light curve at 22 GHz is markedly different
in shape to the light curves at higher frequencies.

The behaviour is thus seen to be complex, particularly after 1990, when a
number of flares, probably about 5, occur over the space of 4 years. These can
be readily distinguished at higher frequencies, but merge together in the radio
regime as a result of the increased time constant.

\subsection{Quiescent spectrum}
\label{plover}
\begin{figure*}
\vspace*{11.7cm}
\caption[dum]{Quiescent spectrum for 3C\,279. See text for details.}
\label{curlew}
\end{figure*}
Investigations of
flaring behaviour of blazars tend to concentrate on modelling epochs
when the flux is significantly raised above some base level.
This base level can be identified with the emission from some underlying jet
along which shocks (identified with the flares)
are propagating (MG85), and is generally assumed to be
quasi-steady in that it is much less variable than the flares which are
superimposed upon it (but see van der Walt 1993). In a previous paper
(Robson \etal 1993) we were able to identify, for 3C\,273,
an extended period of low flux
with this quiescent emission (henceforth we use the word quiescent to refer
to the underlying jet).

\begin{figure*}
\vbox to220mm{\vfil Figure to go here. \vfil}
\caption[dum]{Sequence of spectra for 3C\,279.}
\label{puffin}
\end{figure*}
\begin{figure*}
\vbox to220mm{\vfil Figure continued. \vfil}
\contcaption{}
\end{figure*}
\begin{figure*}
\vspace*{9.5cm}
\contcaption{}
\end{figure*}
It can be seen from Fig.\ \ref{petrel} that during the periods 1986--1994 the
nearest thing to a quiescent state occurred in 1989--1990 which separated
the two large flares of 1988 and 1991. During this period
the flux was largely stable at 375 and 270 GHz, somewhat noisy at 230 GHz,
and declining slightly at 150 GHz. We bin the data over some
arbitrarily selected period and adopt the weighted mean of the data
in the bin as the average quiescent flux. This procedure can be used
for both the sub-mm and IR regions. The
monitoring in the infrared during this time was rather sparse and also rather
noisy (indicating a fair amount of activity in the source, even during
a low state. This is reflected in the noise at 230 and 150 GHz.) This activity
is to be contrasted with the extended period of low flux seen in the infrared
during 1983--early 1986 (Gear \etal 1985, 1986; Brown \etal 1989a).
The flux levels during 1983--1986 are typically within 1 mJy of the lowest
flux levels seen in 1990, and for this reason it is probably justified to
use the low values taken from Brown \etal (1989a) to estimate
the infrared flux during our nominal quiescent period.

In summary, the IR data were binned from 1983 January 11 to 1986 February 24,
whilst the 375--150 GHz data were binned from 1989 June 20 to 1990 April 7.
For the 90--22 GHz data, selected periods were used when the flux seemed
to have attained its lowest level (1986 January 17 to 1986 April 8).
The resulting quiescent spectrum is shown in Fig.\ \ref{curlew}.
The spectral index, $\alpha$ (defined as $F_{\nu} \propto \nu^{\alpha}$
where $F_{\nu}$ is the flux and $\nu$ frequency)
is calculated for the millimetre (150--375 GHz)
and near-IR by a weighted $\chi^2$
fit. The mm index is found to be $-0.35 \pm 0.02$ whilst the near IR
has a slope of $-1.68 \pm 0.03$. An indication of the quality of the fit is
given by the parameter, $q$, which is required to be above some tolerance
level. $q$ is defined as the probability of chance occurrence, and the
tolerance level is taken as 0.01 (Press \etal 1992). Fig.\ \ref{curlew}
indicates that both fits are acceptable and
the millimetre fit particularly so.
It is to be noted that, in contrast with the case of 3C\,273
(Robson \etal 1993) there appears to be no excess thermal emission over the
non-thermal IR power law (the upper limit to the 10.0 $\mu$m flux is
entirely consistent with the extrapolated near-IR power law. It is therefore
to be assumed that the relative contribution of dust to synchrotron emission
in the near- to mid-IR is much less in 3C\,279 than in 3C\,273.
The values assumed for the quiescent flux are given in Table~\ref{seagull}.
\begin{table}
\centering
\caption[dum]{The quiescent flux density values. Errors are given by the
figures in the brackets.}
\def\baselinestretch{1} \large \normalsize
\begin{tabular}{cc} \hline
Frequency & Flux (Uncertainty) \\
& (mJy) \\
\multicolumn{2}{c}{\mbox{ }} \\
J &   1.82 (0.03) \\
H &   3.05 (0.05) \\
K &   4.66 (0.09) \\
L &  10.3 (0.3) \mbox{ } \\
N & $<75$ \\       \hline
(GHz) & (Jy) \\
\multicolumn{2}{c}{\mbox{ }} \\
300 &   4.01 (0.40) \\
375 &   4.70 (0.06) \\
270 &   5.23 (0.08) \\
230 &   5.51 (0.08) \\
150 &   6.45 (0.10) \\
90  &   7.42 (0.37) \\
37  &   8.32 (0.18) \\
22  &   8.55 (0.15) \\ \hline
\end{tabular}
\label{seagull}
\end{table}

\section{Spectral Evolution}
\label{grebe}
There are two distinct approaches to investigating the spectral behaviour of
a blazar in the course of flaring. The first is to examine a sequence of
complete spectra (log flux vs.\ log frequency) and to compare some features
(\eg turn-over, maximum flux) of successive spectra with the
predictions of a given model. The second approach, investigation of the
dependence of spectral index on flux, will be attempted in
section~\ref{rockhopper}.

Previous theoretical discussions have in part assumed that the flare emission
can be described as being from a homogeneous self-absorbed synchrotron
emitting slab
(see MG85, Valtaoja \etal 1988). To check this assumption, and to compare
the observations directly with existing models,
we need to be able to separate the underlying flux
from the flare component. In previous work this has been done
by subtraction of an identified quiescent flux value from the
total observed flux (MG85, Robson \etal 1993; see also Stevens \etal 1995).
In this section we show a sequence of `snapshot' spectra and discuss the
evolution in terms of existing models.

Fig.\ \ref{puffin} shows a sequence of spectra. The requirement for inclusion
was complete, simultaneous 375--150 GHz coverage.
The 90--22 GHz data were
linearly interpolated from adjacent points.
The spectra are a composite of flare flux and underlying flux, and so
homogeneous slab curves will be generally poor fits to the spectra.

\begin{figure*}
\vspace*{9.5cm}
\caption[dum]{Quiescent flux subtracted spectra for 3C\,279.}
\label{heron}
\end{figure*}
\begin{figure*}
\vspace*{15cm}
\caption[dum]{Pre-flare subtracted spectra for 3C\,279.}
\label{avocet}
\end{figure*}

Some trends in turn-over frequency and peak flux are discernible. The first
two panels of Fig.\ \ref{puffin}
show a decay of an old flare at 90 GHz with the clear propagation of
peak flux to lower frequencies. At the beginning of 1990 the flux increases at
higher frequencies and the peak flux is at about 90 GHz. In 1991
a propagation of the peak to lower frequencies is seen,
followed by increased emission at 90 GHz after which the process repeats.
This behaviour occurs throughout the observing epochs. The rise of flux at
higher frequencies precedes the propagation to 37 and 22 GHz (similar behaviour
was seen for 3C\,273 in Robson \etal 1993). It should be noted that, in
general, a power-law description of the 375--150 GHz data is appropriate.
However, direct interpretation of the evolution of the flare
in terms of \eg the model of MG85
(by superposition of a synchrotron spectrum onto a quasi-steady underlying
component) is not possible unless we can isolate the flaring component
directly. We discuss methods of achieving this below.

In common with the approach of MG85, Valtoaja \etal (1988),
and Robson \etal (1993) we can identify the underlying flux with the
quiescent flux (Table \ref{seagull}) and subtract this from the total flux
in order to isolate the flare flux.
Note that we require that each data point remains at least a
$3\sigma$ detection after subtraction.
Examples are shown in Fig.\ \ref{heron}, where
we have fitted each spectrum with a homogeneous synchrotron slab
spectrum with variable turn-over flux, frequency and optically thin slope.
It is clear that the resulting spectra are not usually well described by
emission from a homogeneous synchrotron slab.

There are a number of reasons why this should be so.
That we fail to obtain the homogeneous synchrotron
form for the flare spectrum may question, for example,
the homogeneity of the emitting region. Inhomogeneities
will tend to flatten the spectrum at frequencies below the
turn-over, but will not make a significant difference to the peak flux and
frequency. The assumption of homogeneity is reasonable
during the early stages when the shock is thin.

In van der Walt (1993), detailed calculations were presented which showed that
for an adiabatic, expanding shock, the emission from the unshocked jet is
variable, and can be
substantially less than the quiescent levels. This leads to an underestimate of
the flaring component when it is calculated by subtraction. However,
the shock in the MG85 shock model is thin in comparison with
that assumed by van der Walt (1993), the dominant loss mechanisms being
Compton and synchrotron cooling. We can assume, therefore, that
the unshocked jet emission is reasonably constant during the lifetime of
the shock, at least for frequencies $\gta 90$ GHz.

In a recent paper, Sincell \& Krolik (1994) have suggested that induced
Compton scattering of photons by medium energy relativistic electrons
may provide a dominant opacity mechanism in compact
radio sources. This would result in a number of modifications to the
output emission, and in particular would produce a flattening of
the optically thick (longer wavelength) end of the spectrum.
This could contribute to the appearance of the quiescent spectrum, but
it is less certain that the shocked emission would be modified to any
great extent. Induced Compton scattering is expected to be important if
$\sigma_{\rm T} N_{\rm c} k T_{\rm b}/ m_{\rm e} c^2  > 1 $
(Sincell \& Krolik 1994), where $T_{\rm b}$ is the brightness temperature
and $N_{\rm c}$ is the column density of the scattering electrons.
Setting $T_{\rm b}$ to be $2 \times 10^{11}$ K
(Sincell \& Krolik 1994), we have
$N_{\rm c} \gta 5 \times 10^{22}$ cm$^{-2}$
for induced Compton scattering to be important. The relative thinness
of the shocked region (of order $10^{-3}$ pc during the early stages,
\eg MG85, but c.f.\ Valtoaja \& Ter\"{a}sranta 1994) suggests that
this inequality may not be satisfied for plausible values of the
electron number density, a quantity which is highly sensitive to the
model parameters (not least the lower limit to the electron energy
distribution.) The issue is uncertain, but optically
thick spectral indices of $\sim 2.5$ have been found for 0420$-$014
(Stevens \etal 1995) and 3C\,345 (Stevens \etalc, in preparation) which
indicates that induced Compton scattering cannot at least
be important for all blazars.

A further likely problem is the long term persistence of
slowly decaying flares which add to the flux at lower frequencies.
Inspection of the light curves of Fig.\ \ref{petrel} shows that neither the
22 GHz nor the 37 GHz emission decays to a stable base level. Consequently,
the quiescent fluxes shown in Fig.\ \ref{curlew} correspond to the lowest
flux values seen in Fig.\ \ref{petrel}. We cannot know if these correspond to
anything more than local minima. Thus, any single
observation will generally consist of the sum of
the quiescent flux, flux from the (most recent) flare (which
we are seeking to isolate),
and flux from an unidentified number of residual, decaying flares.
These additional components may be resolved by VLBI,  if available,
(but not easily at $\nu \gta 90$ GHz).
The point is that subtraction of the lowest available flux epoch
(Table~\ref{seagull}) will not remove these residual
components. The net result is an overestimate of the flaring flux at
37, 22 GHz. This is particularly important for post 1992 when the
activity of the source increased dramatically. In particular, the excess
flux at these low frequencies can obscure the turn-over and thus hinder
comparison with theory.

We suggest the following procedure to compensate for these additional
components. Instead of subtracting the quiescent spectrum, we can identify
a pre-flare component (\ie a spectrum from a time immediately before the
flare was seen at higher frequencies). This will be largely identical to
the quiescent spectrum for the 150--375 GHz emission, but significantly above
the quiescent levels between 22 and 90 GHz. If we assume that the rate of
increase of the flare flux is greater than the rate of decrease of the flux
of the previous components, then, at least in the early stages, the pre-flare
spectrum will be a reasonable approximation to a steady underlying component.
It should then be possible to isolate the flare by subtraction of the pre-flare
spectrum.

Fig.\ \ref{avocet} shows the results of subtracting selected pre-flare
spectra.
The fitted optically thin slopes lie from $-0.62$ to $-0.94$, with a
mean of $-0.75$. These values are steeper than the quiescent slope of
$\sim -0.4$, perhaps indicative of radiation losses.
The first panel of Fig.\ \ref{avocet} shows a spectrum from early 1989.
This corresponds to a time after the large but under-sampled flare of 1988,
and during the decay of the subsequent smaller event of early 1989 (but
before our notional quiescent epoch of section~\ref{plover}).
As the light curves in the millimetre are so patchy before about 1989,
the pre-flare which has been subtracted is from 1990 January 17. This
has similar flux levels and better spectral coverage than the actual
pre-flare epochs of early 1988.
The remaining panels have had the pre-flare of 1993 January 19 subtracted.
We note that this procedure works best for large amplitude flares (to minimize
the effects of noise), and that this problem is more acute when subtracting
a pre-flare spectrum which has a higher flux than the quiescent spectrum.

The best homogeneous synchrotron curve
in Fig.\ \ref{avocet} was obtained
for an early flare which was relatively isolated, that is, surrounded by
epochs of quiescent flux (above 90 GHz).
For these flares, residual 22 and 37 GHz
emission was successfully removed by subtraction of a pre-flare. After 1992,
an epoch of densely packed flaring occurred, and subtraction of the pre-flare
is less effective at isolating the flaring emission during these times. The
22 GHz point tends to be too high, and the spectra can be better fitted
with power-laws broken at 90 GHz (although the reduction in $\chi^2$ is
marginal). It should be noted, however, that the excess of the 22 GHz point
over the synchrotron curve is in all cases less than about 1.5 Jy.
If the flaring flux is overestimated because we neglect to
remove additional components, but underestimated because the quiescent flux
is too high, then an excess of $\sim 1$ Jy at 22 GHz over the expected
synchrotron form means the magnitudes of the two effects must be nearly
equal. This is not impossible, (but rather coincidental) and suggests that
the both effects are rather small.

\section{Variations in spectral index}
\label{rockhopper}
We now consider the variation of spectral index in
the IR, sub-mm and radio regimes. Straight line fits to data points
may mask any inherent curvature of the frequency regime
fitted but enable spectral trends to be more easily identified.

\subsection{Infrared spectral variations}
\begin{figure}
\vspace*{8.0cm}
\caption[dum]{Near-IR spectral index against log J-band flux.}
\label{guillemot}
\end{figure}
Fig.\ \ref{guillemot} shows log J-band flux against near-IR spectral slope
calculated from a weighted fit to
all the simultaneous \mbox{J--L$^{\prime}$} flux values from
1986 January onwards. The errors in the ordinate are typically of the order
of the symbol size and so are not shown for clarity.
The relative paucity of quiescent points means
that low flux level trends are difficult to identify
(c.f.\ Robson \etal 1993).
However, a Spearman rank-order correlation test indicates
a $>99.99$ per cent confidence level of a correlation in the sense
of flatter slope at higher flux.
These results are in agreement with those of Gear, Robson \& Brown (1986),
Brown \etal (1989b) for well
sampled sources such as OJ\,287, and are
interpreted as an injection of high energy electrons with a
subsequent rapid decay due to energy losses causing an abrupt steepening of
the spectrum. Note that, in Brown \etal (1989b), 3C\,279 itself showed no
evidence for a correlation as discussed above. It is probable that this was
due to the low number of data points available, and more importantly to the
fact that 3C\,279 was in quiescence at the time. This agrees with the
interpretation of the effect as a phenomenon associated with flaring.

\newpage
\subsection{375--150 GHz spectral variations}
\label{kestrel}
\begin{figure}
\vspace*{8.0cm}
\caption[dum]{Millimetre spectral index against log 270 GHz flux.}
\label{albatross}
\end{figure}
In BLZ5 it was found that there was no strong tendency for the
millimetre--sub-mm spectral
index to either steepen or flatten with increasing flux. We repeat this result
here for a slightly enlarged data set (see Fig.\ \ref{albatross}).
The spectral index shown here
is again a weighted least-squares ($\chi^2$) fit
to simultaneous 375--150 GHz data points; fits below a certain
tolerance threshold ($q=0.01$) were ignored. A Spearman rank-order
test shows no evidence of a correlation between flux and spectral slope
(the confidence level for such a correlation is only
$\sim 15$ per cent), in agreement with the findings of BLZ5.
Other sources exhibiting strong flaring behaviour (\eg 3C\,345,
3C\,273, 0235$+$164, see BLZ5) show a strong correlation of
flattening spectral slope with increased flux. This has been variously
explained as an effect of the passage of the self-absorption turn-over
through the viewing window, or again as the effect of radiative losses upon
a flatter, recently injected electron population. If the former is correct,
and if 3C\,279 has a particularly low frequency turn-over when flaring,
then we would expect any correlation to be weak. However, analysis of the
flares in BLZ5 revealed that the absorption turn-overs of other sources
(\eg 0235$+$164) are at even lower frequencies, which argues against this
interpretation. In the latter case, it
might be possible that such radiation losses are not as important for 3C\,279.
However, we suggest that undersampling of the large flare is the most probable
explanation for the absence of any strong correlation.
We note from \mbox{Fig.\ \ref{albatross}}
that the maximum 270 GHz flux used
was around 14 Jy; the flare was fairly under-sampled above this flux level in
terms of the simultaneous 375--150 GHz flux measurements necessary for
calculation of the spectral slope. It may be, therefore, that the expected
trend has been masked by the relatively low flux data used in
Fig.\ \ref{albatross}. Further data at high flaring epochs are needed to
settle this point.

\subsection{22--90 GHz spectral variations}
The temporal lag between flux maxima is seen to become measurable between
frequencies below 90 GHz, and it is at these
frequencies that the flare amplitudes begin to plateau off or decline, in
agreement with the model of MG85 (see BLZ5). We now contrast the spectral
behaviour of 3C\,279 in the optically thin (millimetre) regime
with its optically thick behaviour.

\begin{figure*}
\vbox to220mm{\vfil Figure to go here. \vfil}
\caption[dum]{22--90 GHz spectral index against log 90 GHz flux.
Linear fits to the rise (circles) and fall (squares)
of each flare are shown.}
\label{penguin}
\end{figure*}
Fig.\ \ref{penguin} shows log 90 GHz flux against the two-point (straight-line)
22--90 GHz spectral index for 3C\,279 from November 1989 onwards.
The data train has been divided into sections
corresponding to the rise and fall of individual flares at 90 GHz
(panels a--c; see Table~\ref{peewit}).
Since the
two sets of observations were not simultaneous,
linear interpolation was performed on the better sampled of the two data
sets, namely that for 22 GHz.
Flares are distinguished on the basis of
clear changes of gradient (usually the onset of the new flare is marked
by a steepening of the rate of change of flux); flares are sub-divided
with the point of maximum flux being assigned to the
rise phase. The exceptions are for the flare starting late 1989 (the
decay of which was obscured by the fast rise of the following flare), and
the rise of the flare of late 1993 which was treated as one with the fall of
the previous flare due to the lack of data points.
Fig.\ \ref{penguin} shows that, for each section,
there is a clear linear correlation between log 90 GHz flux and spectral
index. The lines are given by straight line fits to the data taking
into account the errors in both flux and spectral index (\eg Press \etal 1992).
\begin{table}
\centering
\caption[dum]{Slopes of the linear fit to each
division of the data train shown in Fig.\ \ref{penguin}.}
\def\baselinestretch{1} \large \normalsize
\begin{tabular}{cccc} \hline
Panel & Dates & Flare & $\aldo$ ($\sigma$)\\
(a)& 1989 Oct 31--1990 Aug 15 & Rise & 1.72 (0.14)\\
(b)& 1990 Aug 27--1990 Dec 31 & Rise & 1.34 (0.24)\\
(b)& 1991 Jan 11--1991 May 29 & Fall & 1.99 (0.28)\\
(c)& 1992 Dec 30--1993 May 01 & Rise & 1.03 (0.31)\\
(c)& 1993 Jun 01--1993 Dec 29 & Fall & 1.64 (0.64)\\ \hline
\end{tabular}
\label{peewit}
\end{table}

We note that i) the slopes ($d \alpha / d \log S_{90} \equiv \aldo$)
of each of the fitted straight lines are roughly the same, ranging from
1 to 2; and ii) the decay portion of each flare has a significantly steeper
slope than the rise part. We attempt to provide a theoretical
basis for these results in the next section.

\section{Discussion}
\subsection{Analytic treatment of the 22--90 GHz spectral index flux
correlation}
\label{dodo}
Consider the 22--90 GHz spectral index (henceforth $\alpha$) defined as
follows:
\beq \alpha = k \log (S_{90}/S_{22}) \label{bittern} \eeq
where $k = 1/ \log (90/22)$, and $S_i$ represents the total observed flux at
frequency $i$, which will normally be the sum of flare ($F_i$) and
base ($B_i$) components.
The base or underlying component will in turn be the sum of the quasi-steady
quiescent flux and any residual decaying flux from previous flaring events.

We adopt a simplified view of the flaring component by assuming it to be two
power-laws intersecting at some turn-over
frequency $\nu_{\rm m}$ with an optically thin
slope $a$ for $\nu > \nu_{\rm m}$ and an optically thick index $b$ for
$\nu < \nu_{\rm m}$ ($a, b \neq 0$). The ratio of the two flare fluxes
(henceforth $F_{90}$, $F_{22}$
respectively) is a function of the parameters $a$, $b$ and the position of the
turn-over $\nu_{\rm m}$. If $\nu_{\rm m} > 90$ GHz, then
\beq F_{90}/F_{22} = (90/22)^{a}. \eeq
During the early stages of the flare,
the flare flux will dominate the base flux at 90 GHz,
but not at 22 GHz (\ie $F_{90} \gg B_{90}$, and $S_{22} \simeq B_{22}$).
If this is the case then $S_{22}$ will be roughly
constant, and hence $\aldo = k$ from equation~(\ref{bittern}).

The turn-over frequency of the flare will evolve to lower frequencies, until
at some point $22<\nu_{\rm m}<90 $ GHz. If we
denote the flare flux at $\nu_{\rm m}$ by $F_{\rm m}$, then we have
$F_{\rm m}/F_{22} = (\nu_{\rm m}/22)^b$ and
$F_{90}/F_{\rm m} = (90/\nu_{\rm m})^a$.
Thus
\beq F_{22} = F_{\rm m}^{1-b/a} F_{90}^{b/a} (22/90)^b  \mbox{.}
\label{tern} \eeq
The flare turn-over flux ($F_{\rm m}$) has
predicted evolutionary paths
in flux/frequency space according to MG85. Specifically,
$\nu_{\rm m}= \zeta F_{\rm m}^{1/c}$
where $\zeta$ is a constant of proportionality and $c$ is a
constant which depends on the phase of the flare evolution, and also on $a$
and $b$. Thus for non-zero $c$
\beq F_{\rm m} = F_{90}^{c/(c-a)} (\zeta/90)^{ca/(c-a)}
\mbox{.} \label{cormorant} \eeq
During the later stages of the flare's lifetime, we make the assumption that
the flare flux dominates any underlying
component at both 22 and 90 GHz (\ie $F_i \gg B_i$).
Thus, substituting into equation~(\ref{bittern})  from
equations~(\ref{tern}) and~(\ref{cormorant}), we can write
\beq \alpha = k \left( \frac{b-a}{c-a} \right) \{ \log(S_{90})
+ c \log(\zeta/90) \} + k b \log(90/22)
\label{shag} \eeq
which is linear in $\log(S_{90})$.
For the
synchrotron phase appropriate to the canonical value of
$a=-0.75$ we have $c=0$, and $F_m$ is constant. In this case the equivalent
formula to equation~(\ref{shag}) is
\beq \alpha = k(1-b/a) \log(S_{90}) -k \log(F_{\rm m}^{1-b/a}(22/90)^b)
\label{pippit} \eeq
which is again linear in $\log(S_{90})$.

Eventually $\nu_{\rm m} < 22$ GHz, and $F_{90}/F_{22} = (90/22)^{b}$, and
hence $\alpha$ will be roughly constant before returning to the quiescent
value.

The early stage of the flare (when the 22 GHz flux is assumed constant
and probably $\nu_{\rm m} > 90$ GHz)
has a slope ($\aldo$) of $k$, equal to 1.63. Subsequent stages (as the
self-absorption turn-over evolves to shorter frequencies) have values of
$\aldo$ critically dependent on the nature of the flare, specifically
the values of the optically thin and thick spectral indices $a$ and $b$.
The value of $a$
in turn determines the value of $c$ for the later (synchrotron and adiabatic)
phases of the flare (see MG85 for derivation of the appropriate formulae).
The usual value of $b$ for an homogeneous synchrotron self-absorption
turn-over is $b=2.5$. If the optically thin spectral index
has the canonical value of $a=-0.75$, then the
synchrotron phase has $c=0$ whilst the decay phase has $c=0.59$.
Using equations~(\ref{pippit}) and~(\ref{shag}) this gives values of $\aldo$
of 7.1 and 4.0 respectively. Both are considerably steeper than observed.
Whilst the values of the thin spectral index seen in Section~\ref{kestrel} are
typically around $-0.5$, slightly flatter than canonical, Fig.\ \ref{avocet}
suggests that the flare spectrum has a steeper optically thin slope, perhaps
as steep as $-0.9$. It can be seen from equation~(\ref{shag}) that, for a
given $b$, $\aldo$ flattens as $a$ decreases. This would agree
with the slopes in Fig.\ \ref{penguin}.

The behaviour of the flare according to this analytic treatment
is summarized in Fig.\ \ref{ducky}. The evolution of the
self-absorption turn-over is shown by the solid lines. Different phases
of the evolution have different values of $\aldo$, and are delimited by
vertical lines. The numbers are the values of $\aldo$ appropriate
to $a=-0.75$ and $b=2.5$ (with corresponding values of $c$).
\begin{figure}
\vspace*{8.cm}
\caption[dum]{Schematic summary of one possible example of the flaring
behaviour treated analytically in the text. Both flux and frequency are
printed on a logarithmic scale; the arrow indicates the direction
of evolution of the flare; the numbers are the values of $\aldo$
expected.}
\label{ducky}
\end{figure}

\subsection{Qualitative discussion}
It is argued below that the values of $\aldo$ will in general be flatter than
calculated above. We suggest a number of
possible explanations. For instance: the flare
spectrum is not well described by two power-laws; such an approximation is
bound to be rather poor, particularly near the absorption turn-over, where
there is considerable curvature. Inclusion of
this factor will tend to reduce the value of $b$ and increase
that of $a$; \ie to flatten both indices (see section~\ref{auk}).
Furthermore, the neglect of the underlying base component is almost certainly
an oversimplification, particularly when considering
the increased decay time at lower
frequencies. The effects of residual emission are likely to increase the
underlying 22 GHz flux. If the flare flux does not dominate the base flux at
this frequency, then the effect will be to reduce the effective value of $b$.
Inhomogeneities in the  emitting region will have the same effect (MG85).
If $b$ is lower than 2.5 then the decay parts of the flares will have
lower $\aldo$ than expected, as will the rising portions.
This would be expected to occur
during epochs of densely packed flaring, \ie late 1990 onwards,
which is exactly as seen (see also Fig.\ \ref{avocet}).

One of the key features of the observations (Fig.\ \ref{penguin})) is that
the rise to flare maximum has a lower value of $\aldo$ than the decay.
This feature was seen to be common to all the flares observed, and
is successfully accounted for by the treatment of section~\ref{dodo}.
If the plateau phase of
evolution is either generally short lived, or narrow in terms of frequency
range, then the chance of it being observed is probably rather low.
The resulting behaviour would be exactly as seen in Fig.\ \ref{penguin}:
two straight lines appearing to intersect,
with the upper line corresponding to the rise of the flare and the
steeper, lower one to the decay. Note that we see no strong evidence for
an extended plateau phase in the previous section, nor was there any evidence
seen in BLZ5. This supports the conclusion that the plateau phase is rather
short, as might be expected from the dominance of gamma ray luminosity to
synchrotron emission, assuming the former to be produced by the
inverse Compton process.
Quantitative estimates for the extent of the plateau phase are given below.

\subsection{Modelling of the behaviour}
\label{auk}
There is scope for modelling of
the passage of the flare in terms of the change in $\alpha$ as
a simple synchrotron curve moves through the 22, 90 GHz window.
To do this we can adopt a simple parametric description of the shock model
of MG85 as described above. In this version of the generalized
shock model the behaviour of the flares is parameterized by fixing the
evolutionary tracks of the self-absorption turn-over in flux/frequency space,
and assuming that the spectrum at any given time can be described by a
homogeneous slab synchrotron curve. It is then necessary to adopt a suitable
function for the dependence of turn-over frequency with time (typically
$\nu_{\rm m} \propto t^{\delta}$, where $\delta$ depends on the evolutionary
phase). Flare light curves can thus be generated at any desired frequency.
Evolutionary tracks for the turn-over
appropriate to this value are given by $S_{\rm m} \propto \nu_{\rm m}^c$,
where $c$ has the values appropriate to an optically thin spectral index of
$-0.75$ (\ie $c=-2.5, 0, 0.59$ for the three phases).
The transitions between growth/plateau phases, and
plateau/decay phases are chosen as 90 and 60 GHz, respectively. The 90 and
22 GHz light curves thus generated are shown in Fig.\ \ref{corncrake}.

The observed duration
is a function not only of the frequency width of the phase (here set at
30 GHz),
but also of the relationship between turn-over frequency and time (i.e how long
the flare spends on that phase), and also of the sampling frequency.
The delimiting values for the synchrotron phase (90 and 60 GHz) were chosen
such that the observed duration of that phase was short. The values
chosen for the extent of this phase are probably not an unreasonable estimate,
and are consistent with the amplitude analysis of BLZ5. The time
spent in the synchrotron phase is probably of order weeks rather than months.
This is comparable to the length of the Compton phase, which is similarly
expected to be rapid (MG85).
\begin{figure}
\vspace*{8.cm}
\caption[dum]{Simulated 90 and 22 GHz light curves (solid and dotted lines
respectively).}
\label{corncrake}
\end{figure}

\begin{figure*}
\vspace*{11.cm}
\caption[dum]{Simulated 22--90 GHz spectral index against log 90 GHz flux.
See text for an explanation of the symbols.}
\label{shrike}
\end{figure*}
Fig.\ \ref{shrike} shows a plot of simulated 22--90 GHz spectral
index varying with log 90 GHz flux. Note the resemblance between
this figure and Fig.\ \ref{penguin}.
The simulation is produced in the
manner described above with the addition of irregular sampling (using a
uniform deviate) and superposition
of the quiescent spectrum of Table~\ref{seagull}. We also impose a random
scatter on each data point, the amount of scatter chosen using
a Gaussian deviate with standard deviation of 5 per cent of the flux.
Points for which $\nu_{\rm m} > 90$ GHz are
shown by a cross, whilst those for which $\nu_{\rm m} < 22 $ GHz are shown
by a square. Those points for which $\nu_{\rm m} \in [22,90]$ GHz are marked
by a triangle if the turn-over lay on the synchrotron phase, and by a circle
if it lay on the adiabatic phase. The synchrotron phase is
rapid, and hence few triangles are seen. Note also that the Compton phase is
also fairly short lived in comparison with the later decay phases, and so
there are fewer crosses than circles. This agrees with the real data depicted
in Fig.\ \ref{penguin}.

The crosses, corresponding to the early Compton phase, are well described
by a straight line with a slope of $\sim 1.4$. The
points marked by circles also lie on a straight line (as expected from
the analytic discussion above) and their slope is found by least squares
fitting to be $\sim 2.5$. This is slightly flatter than that expected from
equation~(\ref{shag}) which probably reflects the effect of the
curvature inherent in the synchrotron spectrum and the
contribution from the underlying flux. A value of $\sim 2.5$ is,
however, steeper than that seen  in Fig.\ \ref{penguin}.

According to the treatment of section~\ref{dodo}, when the flare peaks at
22 GHz, (corresponding to $\nu_{\rm m} < 22$ GHz), then
the value of $\alpha$ is roughly
constant before returning to the quiescent value. This phase is shown by the
squares in Fig.\ \ref{shrike}. However, in Fig.\ \ref{penguin} we see no
similar behaviour.
This may well be due to the fact that
Fig.\ \ref{penguin} has points from a number of successive flares, rather than
an isolated example as modelled above. In other words, this may yet be another
example of the difficulty of disentangling the flux from many flares: the
90 GHz emission probably contains a significant contribution from newer flares.

\section{Conclusions}
\parindent 0em
The main conclusions of this paper can be summarized as follows:

1) We find a general trend for the maximum flux to
propagate to lower frequencies, which is consistent with a shock expanding
down a jet (MG85, Hughes, Aller \& Aller 1989).
Because of the complexities of the flaring, we cannot critically test
the details of the MG85 model (but c.f.\ Stevens \etal 1995). We suspect that
the behaviour is not entirely
consistent in detail, perhaps because of difficulties is disentangling the
flaring emission. Alternatively, the model itself may need some modification
(see \eg Marscher, Gear \& Travis 1992).

2) We have found examples for which the quiescent or
pre-flare subtracted spectra resembled the homogeneous slab synchrotron form.
Deviations from this form at low frequencies
may be accounted for by inhomogeneities in the emitting
region (MG85).

3) The infrared spectral index is shown to be strongly correlated with
log J-band flux. The confirms earlier results for OJ\,287 (Gear, Robson \&
Brown 1986, Brown \etal 1989b),
and is interpreted as radiation losses (\eg Kardashev 1962).

4) No such effect is seen for weighted 375--150 GHz spectral indices which
are uncorrelated with log 270 GHz flux, in contrast with other sources
(\eg 3C\,273, BLZ5). This may be due to undersampling of the peaks of
the flares.

5) We see strong linear correlations between two-point 22--90 GHz spectral
indices and log 90 GHz flux. This trend exists separately for
the rise and fall of each flare isolated. The slopes of the linear trends
can be found by least squares fitting, and are seen to be steeper
on the decay of a flare than on the rise.

6) A simple analytic treatment is presented which can explain these linear
relations in terms of the evolution of the flare according to the MG85 model.
In particular, analytic expressions are given which explain the steepening
of the decay of the flare in comparison to the rise. Numerical simulations
of the same MG85 model similarly reproduce the observed trends, and
show better agreement with the observed values of $\aldo$.

\parindent 18pt

3C\,279 is revealed as a complex source, showing many flaring events over
a short period of time. The extreme variability of the source may be a
consequence of the fact that it is so well sampled (as is the case for
3C\,273, see Robson \etal 1993). However, in
BLZ5 it was shown, particularly from
the 37 and 22 GHz data in that paper, that some sources are more dramatic
and variable than others, and 3C\,279 certainly falls into the former
category. We therefore believe that
better temporal sampling of flaring events is needed
for a successful deconstruction of the spectral behaviour of this source.

In particular, in order to perform a proper analysis,
we require an isolated and major flare with a lack of
sub-flaring (Robson 1992).
Weekly or monthly monitoring at a single high frequency may
show when 3C\,279 has reached a stable state and enable identification of
a flare in the
early stages of formation. From that stage regular daily monitoring
can be employed to follow the flare evolution in its entirety.
In particular, to directly observe the Compton phase of
the flare's growth (which is liable to be rapid, as pointed out in MG85),
near- and mid-infrared observations simultaneous with
the millimetre are needed, on daily (or even hourly) monitoring time-scales
(Robson 1992).

\section*{ACKNOWLEDGEMENTS}
The James Clerk Maxwell Telescope is operated by the Royal Observatories
on behalf of the United Kingdom Particle Physics and Astronomy
Research Council (PPARC), the Netherlands Organization for the Advancement
of Pure Research, the Canadian National Research Council (NRC) and the
University of Hawaii.
SJL acknowledges financial support from the UK PPARC,
and the University of Central Lancashire.
JAS acknowledges a research studentship from
the University of Central Lancashire.
We gratefully acknowledge
use of the STARLINK computing facilities at the Centre for Astrophysics.

\bsp
\end{document}